\documentclass[preprintnumbers, floatfix, preprintnumbers, letterpaper, superscriptaddress,nofootinbib, twocolumn]{revtex4}
\pdfoutput=1
\usepackage{graphicx}
\usepackage{microtype}
\usepackage{amsmath}
\usepackage{amssymb}
\usepackage{subfigure}
\usepackage{hyperref}
\usepackage{url}
\usepackage{xcolor}
\usepackage{color}
\usepackage{mathrsfs}
\usepackage{calrsfs}
\usepackage{amsfonts}
\usepackage{latexsym}
\usepackage{ragged2e}
\usepackage{epsfig}
\usepackage{textcomp}

\usepackage{caption}
\DeclareCaptionJustification{justified}{\leftskip=0pt \rightskip=0pt \parfillskip=0pt plus 1fil}
\definecolor{vividviolet}{rgb}{0.62, 0.0, 1.0}
\definecolor{amaranth}{rgb}{0.9, 0.17, 0.31}
\definecolor{palatinateblue}{rgb}{0.15, 0.23, 0.89}
\definecolor{brightpink}{rgb}{1.0, 0.0, 0.5}
\definecolor{cornflowerblue}{rgb}{0.39, 0.58, 0.93}
\definecolor{deepcarminepink}{rgb}{0.94, 0.19, 0.22}
\definecolor{radicalred}{rgb}{1.0, 0.21, 0.37}

\hypersetup{ linktoc=all,
    colorlinks, linkcolor={palatinateblue},
    citecolor={brightpink}, urlcolor={amaranth}
}

\graphicspath{{Images/}}

\graphicspath{{Images/}}

\makeatletter
\def\@fnsymbol#1{\ensuremath{\ifcase#1\or $\textleaf$ \or \ddagger
\else\@ctrerr\fi}}%

\makeatother

\def\sideremark#1{\ifvmode\leavevmode\fi\vadjust{\vbox to0pt{\vss
 \hbox to 0pt{\hskip\hsize\hskip1em
 \vbox{\hsize1.5cm\tiny\raggedright\pretolerance10000
 \noindent #1\hfill}\hss}\vbox to8pt{\vfil}\vss}}}%
\begin{document}

\title{Generalized Uncertainty Principle and White Dwarfs Redux: \\How Cosmological Constant Protects Chandrasekhar Limit}

\author{Yen Chin \surname{Ong}}
\email{ycong@yzu.edu.cn}
\affiliation{Center for Gravitation and Cosmology, College of Physical Science and Technology, \\Yangzhou University, Yangzhou 225009, China}
\affiliation{Nordita, KTH Royal Institute of Technology \& Stockholm University,
Roslagstullsbacken 23, SE-106 91 Stockholm, Sweden}

\author{Yuan \surname{Yao}}
\affiliation{Center for Gravitation and Cosmology, College of Physical Science and Technology, \\Yangzhou University, Yangzhou 225009, China}

\begin{abstract}
It was previously argued that generalized uncertainty principle (GUP) with a positive parameter removes the Chandrasekhar limit. One way to restore the limit is by taking the GUP parameter to be negative. In this work we discuss an alternative method that achieves the same effect: by including a cosmological constant term in the GUP (known as ``extended GUP'' in the literature). We show that an \emph{arbitrarily small but nonzero} cosmological constant can restore the Chandrasekhar limit. We also remark that if the extended GUP is correct, then the existence of white dwarfs gives an upper bound for the cosmological constant, which -- while still large compared to observation -- is approximately 86 orders of magnitude smaller than the natural scale. 
\end{abstract}

\maketitle

\section{Generalized Uncertainty Principle and White Dwarfs}\label{Intro}

The generalized uncertainty principle (GUP) is a quantum gravity inspired correction to the Heisenberg's uncertainty principle, which reads (in the simplest form)
\begin{equation}
\Delta x \Delta p \geqslant \frac{1}{2} \left[\hbar + \frac{\alpha L_p^2 \Delta p^2}{\hbar}\right],
\end{equation}
where $L_p=1.616229\times10^{-35}$ m denotes the Planck length, and $\alpha$ is the GUP parameter typically taken as an $O(1)$ positive number in theoretical calculations, i.e. one expects that the GUP correction becomes important at the Planck scale. GUP is largely heuristically ``derived'' from Gedanken-experiments under a specific quantum gravity theory (such as string theory \cite{5,6,7,8}), or general considerations of gravitational correction to quantum mechanics \cite{9301067, 9305163, 9904025, 9904026}. GUP is useful as a phenomenological approach to study quantum gravitational effects. From phenomenological point of view, the GUP parameter can be treated as a free parameter \emph{a priori}, which can be constrained from experiments \cite{1607.04353, 1610.08549v4, 1804.03620, 1704.02037}; $\alpha$ as large as $10^{34}$ is consistent with the Standard Model of particle physics up to 100 GeV \cite{1607.04353}, while a tunneling current measurement gives $\alpha \leqslant 10^{21}$ \cite{0810.5333}. See also \cite{1804.06389, 1512.07779, 1107.3164}.

It turns out that GUP has a rather drastic effect on white dwarfs. This is somewhat of a surprise, since we do not usually expect GUP correction to be important for scale much above the Planck scale. 
A standard -- though hand-wavy -- method to obtain the behavior of degenerative matter is to consider the uncertainty principle $\Delta x \Delta p \sim \hbar$, and then take $\Delta x \sim n^{-1/3}$, where $n$ is the number density $n=N/V=M/(m_e V)$ of the white dwarf (here modeled as a pure electron star), where $N$ is the total number of electrons, whereas $V, M$ are, respectively, the volume and the total mass of the star, and $m_e$ the electron mass. Then, the total kinetic energy in the non-relativistic case is 
\begin{equation}
E_k = \frac{N \Delta p^2}{2 m_e} \sim \frac{M^{\frac{5}{3}} \hbar^2}{2 m_e^{\frac{8}{3}}R^2},
\end{equation}
where $R$ is the radius of the star. Equating this with the magnitude of the gravitational binding energy $|E_g| \sim GM^2/R$, one obtains the radius as a function of the mass:
\begin{equation}
R \sim \frac{\hbar^2}{2m_e^{\frac{8}{3}}GM^{\frac{1}{3}}}.
\end{equation}
Thus the more massive a white dwarf is, the smaller it becomes. A similar derivation using the relativistic kinetic energy, and assuming that the momentum dominates over the rest mass of the electrons, allows one to obtain the Chandrasekhar limit up to a constant overall factor (see \cite{1804.05176} for details). That is, the ultra-relativistic curve is a line given by $M=M_\text{Ch}$, where $M_\text{Ch}$ is the Chandrasekhar mass. Modulo some numerical constant that cannot be determined from this simple method, it is\footnote{The actual value is somewhat smaller, \begin{equation}M_\text{Ch}=\frac{\omega^0_3 \sqrt{3\pi}}{2}\left(\frac{\hbar c}{G}\right)^{\frac{3}{2}}\frac{1}{(\mu_e m_H)^2} \approx\frac{5.76}{\mu_e^2}\ M_\odot, \notag\end{equation} where $\mu_e$ denotes the mean molecular weight per electron, while $m_H$ the mass of hydrogen atom, and $M_\odot$ the solar mass. The constant coefficient $\omega^0_3 \approx 2.0182$ is obtained via the Lane-Emden equation. See, e.g., Eq.(43) of \cite{chandra}. For $\mu_e=2$, this yields the familiar $M_{\text{Ch}}=1.44 M_\odot$, i.e.  about $2.86 \times 10^{30}$ kg.} 
\begin{equation}
M_\text{Ch} \sim \frac{1}{m_e^2}\left(\frac{\hbar c}{G}\right)^{\frac{3}{2}} \approx 1.24 \times 10^{37} \text{kg}.
\end{equation}

\begin{figure}[!h]
\centering
\includegraphics[width=3.3in]{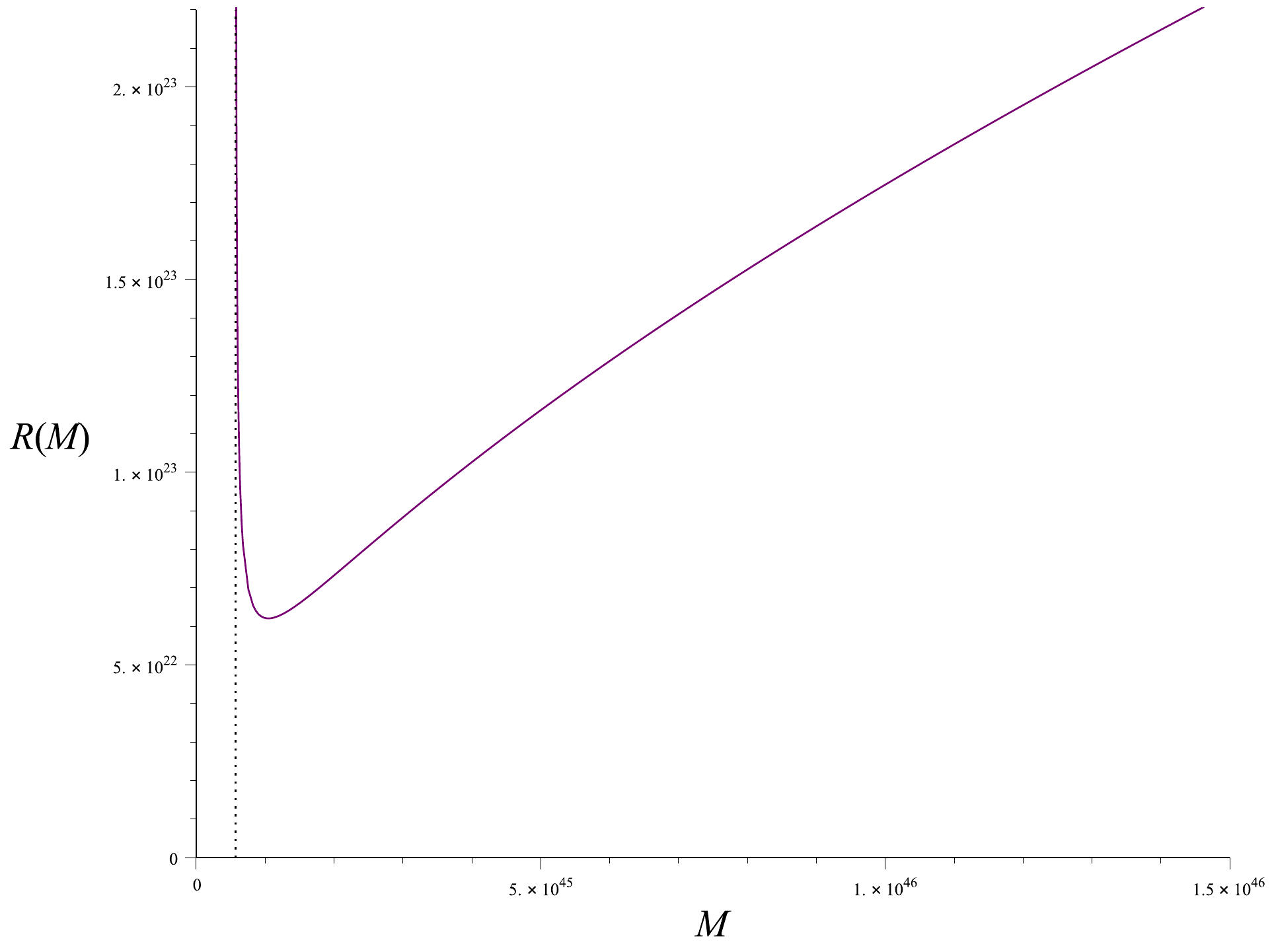}
\caption{The mass-radius relationship of an ultra-relativistic white dwarf with GUP correction. Without GUP correction, it is simply the Chandrasekhar limit shown by the vertical dotted line. However, if $\alpha > 0$, the curve deviating away from the vertical line can eventually ``bounce'' and then grows unbounded in size. Here we have set $G=c=\hbar=1$. We have furthermore set $\alpha=1$ in this example. \label{fig1}} 
\end{figure}

If GUP is used in place of the standard uncertainty principle, then one finds a surprising and disturbing result: the Chandrasekhar limit disappears. More specifically, for the non-relativistic curve, it no longer goes like $R \sim M^{-1/3}$ for large values of the mass, instead $R$ eventually grows with $M$. The ultra-relativistic curve does not yield the Chandrasekhar limit: while the curve $R(M)$ tends to the original line $M=M_\text{Ch}$ as $M \to M_\text{Ch}^+$, it is no longer bounded above when $M$ increases. This means that white dwarfs can in principle gets \emph{arbitrarily large} (for any $M$, there exists a nonzero $R$ that satisfies the equilibrium equation between degenerate pressure and gravity; this is true for non-relativistic case too), consistent with the previous results obtained in \cite{1512.06356v2} using a more rigorous method (see also \cite{1512.04337, 1301.6133,1712.03953}). In Fig.(\ref{fig1}), we show the generic behavior of the ultra-relativistic white dwarf radius-mass relation, in the Planck units, i.e. $G=c=\hbar=1$ (and so $m_e = 4.1854 \times 10^{-23}$).  

Here we summarize the main issues and caveats discussed in \cite{1804.05176}: we find the absence of a Chandrasekhar limit disconcerting, since it seems that an arbitrarily small GUP parameter $\alpha$ can give rise to a huge effect in white dwarfs. Astrophysical observations have indicated that white dwarfs in fact do obey the Chandrasekhar limit (so-called ``super-Chandrasekhar'' white dwarfs are also of $O(M_\text{Ch})$) \cite{0609616, 1106.3510}. There are other possible explanations as to why despite allowed by the GUP, huge white dwarfs (one might call them ``white giants'') are nevertheless absent, notably the interior structures of white dwarfs might have other effects that dominate over that of GUP. Furthermore, one has to be more careful about the $R$-$M$ diagram, since the ``bounce'' and ``growth'' part of the curve might be under the line $R=2GM/c^2$ (i.e. smaller than the Schwarzschild radius) for sufficiently small value of $\alpha$, indicating that these parts are irrelevant since black holes already formed. However, as mentioned in \cite{1804.05176}, whether this happens would require a more detailed study involving a GUP-corrected fully general relativistic description of the white dwarfs to be consistent. In addition, in view of how surprising GUP correction could be, we do not have full confidence that black hole formation criterion is also not modified. Therefore we reserve this possible resolution for future work\footnote{Such a possibility is indeed quite intriguing, since it suggests that much like how general relativity tends to ``censor'' naked singularities and closed timelike curves behind black hole horizons, ``unwanted novel features'' of GUP might also be ``censored''.}. In any case, our present proposal is more straightforward, in the sense that the Chandrasekhar limit is restored explicitly. 

In \cite{1804.05176}, a simple resolution was suggested: take the GUP parameter $\alpha$ to be negative. Such a choice seems rather odd at first glance -- uncertainty reduces as one approaches the Planck scale, so physics becomes more classical -- but is nevertheless in agreement with some models of quantum gravity \cite{0912.2253, 1504.07637}. In this work we propose another possible resolution, which allows $\alpha$ to be positive, by taking into consideration the inclusion of a nonzero cosmological constant in the GUP  (hereinafter, extended generalized uncertainty principle, or ``EGUP''). We further show that the Chandrasekhar limit is protected by an arbitrarily small positive cosmological constant. This is satisfactory since our actual Universe is best described by the concordance $\Lambda$CDM model with exactly such a cosmological constant. As a bonus, this approach yields an upper bound for the cosmological constant, which although is still many order of magnitude higher than the observed value, is considerably smaller than the ``natural'' value from quantum field theoretic estimate. 

\section{The Extended Generalized Uncertainty Principle}

To best way to understand EGUP is to start from Hawking temperature of a black hole. In the case of asymptotically flat Schwarzschild black hole, there is a heuristic way to derive the Hawking temperature up to a constant factor: consider the uncertainty in the position of the Hawking particle being emitted around the black hole, $\Delta x \sim GM/c^2$. The Heisenberg's uncertainty principle gives $\Delta p \sim \hbar/\Delta x=\hbar c^2/GM$, then one relates the temperature to momentum via $E=pc=k_B T$. One thus obtains the Hawking temperature $T \sim \hbar c^3/k_BGM$ \cite{pc}.  

The GUP correction leads to a modification to Hawking temperature, which is derived by repeating the calculation with GUP in place of Heisenberg's uncertainty principle \cite{pc}. However, one notes that the above mentioned heuristic approach to derive the standard Schwarzschild black hole temperature does \emph{not} extend straightforwardly to other more complicated black hole spacetimes. In particular, it does not work for Schwarzschild-de Sitter (dS) or Schwarzschild anti-de Sitter (AdS) black holes, which have in addition to the mass, a length scale $L$, associated with the cosmological constant via the relation $\Lambda=\pm 3/L^2$ in 4-dimensions. 

It was proposed in \cite{0709.2307} that Heisenberg's uncertainty principle can be modified to include the cosmological constant term, to the form called ``extended uncertainty principle'' (EUP),
\begin{equation}\label{EUP}
\Delta x \Delta p \geqslant \frac{1}{2} \left[\hbar + \beta \frac{\hbar(\Delta x)^2}{L^2}\right],
\end{equation}
so that the heuristic method can be applied, \emph{mutatis mutandis}, to derive the correct black hole temperature (up to a constant).
To achieve this, the parameter $\beta$ has to be chosen as $\pm 3$, for AdS and dS, respectively.  So unlike $\alpha$, \emph{the value of $\beta$ is known from theoretical consideration.} Although one can consider freeing up $\beta$ to be arbitrary to explore the parameter space, in this work we fix it to be $-3$ for dS space.
In the limit $\Lambda \to 0$, or equivalently $L \to \infty$, we recover the asymptotically flat result. 

In this work, in view of our interest concerning white dwarfs in the actual Universe, we shall focus on positive cosmological constant only\footnote{It is conceivably possible that \emph{fundamentally} the cosmological constant in our Universe is negative, with the current accelerating expansion caused by other field, such as the quintessence \cite{0403104}. However, the effective cosmological constant is still positive.}. 

Having obtained the EUP, one could then consider the correction term from quantum gravitational effect and therefore ends up with the EGUP

\begin{equation}\label{EGUP}
\Delta x \Delta p \geqslant \frac{1}{2} \left[\hbar + \alpha\frac{L_p^2 (\Delta p)^2}{\hbar} + \beta \frac{\hbar(\Delta x)^2}{L^2} \right].
\end{equation}

The motivation to extend the uncertainty principle in the aforementioned manner is somewhat unconvincing -- it makes the assumption that the heuristic method for deriving Hawking temperature is applicable to Schwarzschild-(A)dS black hole, and then deduce the form of the uncertainty principle that is required to make this works. May one be carrying the heuristic method too far?
There are other ``derivations'', such as considerations of the symmetry of phase space \cite{Bambi}, gedanken experiment on measurement taken in background with nonzero cosmological constant \cite{Bambi}, as well as heuristic derivation using quantum mechanics formulated in de Sitter and anti-de Sitter space \cite{Mignemi, 0911.5695}. (Note, however, that the sign of $\beta$ is argued to be the other way round in \cite{Bambi}.) An attempt to rigorously derive the EUP can be found in \cite{F}. In this work we will take EGUP as given, and explore its consequences to white dwarfs. 

One might object that the uncertainty principle is fundamental and should not be modified depending on background spacetime. However, mathematically, the uncertainty principle concerns Fourier transforms of functions, which is nontrivial on curved manifolds. From the mathematical perspective, we \emph{should} expect some kind of modification to the uncertainty principle due to spacetime curvature (see, e.g., \cite{0306080, 1804.02551}). Since curvature, being a geometric quantity, cannot be transformed away via a coordinate change even at a point, being locally Minkowski does not get rid of the correction term, though it could be miniscule. EUP is therefore not without a basis.

It is worth mentioning at this point that angular momentum $L_z$ and angular coordinate $\phi$,  do not satisfy the usual uncertainty relation $\Delta \phi \Delta L_z \geqslant \hbar/2$ (see, e.g., \cite{judge1, judge2}) due to the periodicity of $\phi$. If we consider a unit circle as a quotient of a line $S^1 \cong \Bbb{R}/\Bbb{Z}$, in a compact spatially 1-dimensional universe, locally an observer does not know the global topology, but nevertheless $\Delta x \Delta p$ would be modified in the same manner and therefore takes the form
\begin{equation}
\Delta x \Delta p \geqslant \frac{\hbar}{2}\left[1-C(\Delta x)^2\right],
\end{equation}
where $C$ is a constant, argued to be $3/\pi^2$ in \cite{judge1, judge2}. Since the (global) spatial section of de Sitter space is a 3-sphere, it is not surprising that EUP takes a similar form (though such a comparison is clearly only suggestive -- the size of $S^3$ in de Sitter space changes with time, but the constant $\beta$ in EUP is fixed. It might be interesting to further investigate this analogy further, however.)

\section{How Cosmological Constant Protects Chandrasekhar Limit}

Since the approach is rather qualitative, we dropped the factor of $1/2$ for convenience in Eq.(\ref{EGUP}) and consider instead 
\begin{equation}\label{EGUP2}
\Delta x \Delta p \sim  \hbar + \alpha\frac{L_p^2 \Delta p^2}{\hbar} + \beta \frac{\hbar(\Delta x)^2}{L^2},
\end{equation}
which yields
\begin{equation}
\Delta p \sim \frac{\hbar \Delta x}{2 \alpha L_p^2} \left[1 \pm \sqrt{1-4\alpha L_p^2\left(\frac{\beta}{L^2}+\frac{1}{\Delta x^2}\right)}\right].
\end{equation}
Following the method in \cite{1804.05176} summarized in Sec.(\ref{Intro}),
we found that, for the non-relativistic case, EGUP-corrected white dwarfs satisfy
\begin{equation}\label{nonrelWD}
M^{\frac{5}{6}} \sim \frac{\hbar R^{\frac{3}{2}}}{2\sqrt{2G}\alpha m_e^{\frac{2}{3}}L_p^2} \left[1 -\sqrt{1 -4\alpha L_p^2\left(\frac{\beta}{L^2} + \frac{M^{\frac{2}{3}}}{m_e^{\frac{2}{3}}R^2}\right)}\right]. 
\end{equation}
The minus sign in front of the square root sign is fixed by requiring that in the large $L$ and small $\alpha$ limit, we recover the standard Heisenberg's uncertainty principle \cite{1804.05176}.
In this work we will not show the plot for this case because it turns out to be essentially indistinguishable from the $\beta=0$ case when $L$ is large. With the value of $L$ corresponds to the cosmological constant in our Universe (see below), the curve turns over at around $M \sim 5\times 10^{79}$, which is way above the Chandrasekhar limit (which is restored, as shown below), and thus this turn-over is irrelevant.

\begin{figure}[!h]
\centering
\includegraphics[width=3.3in]{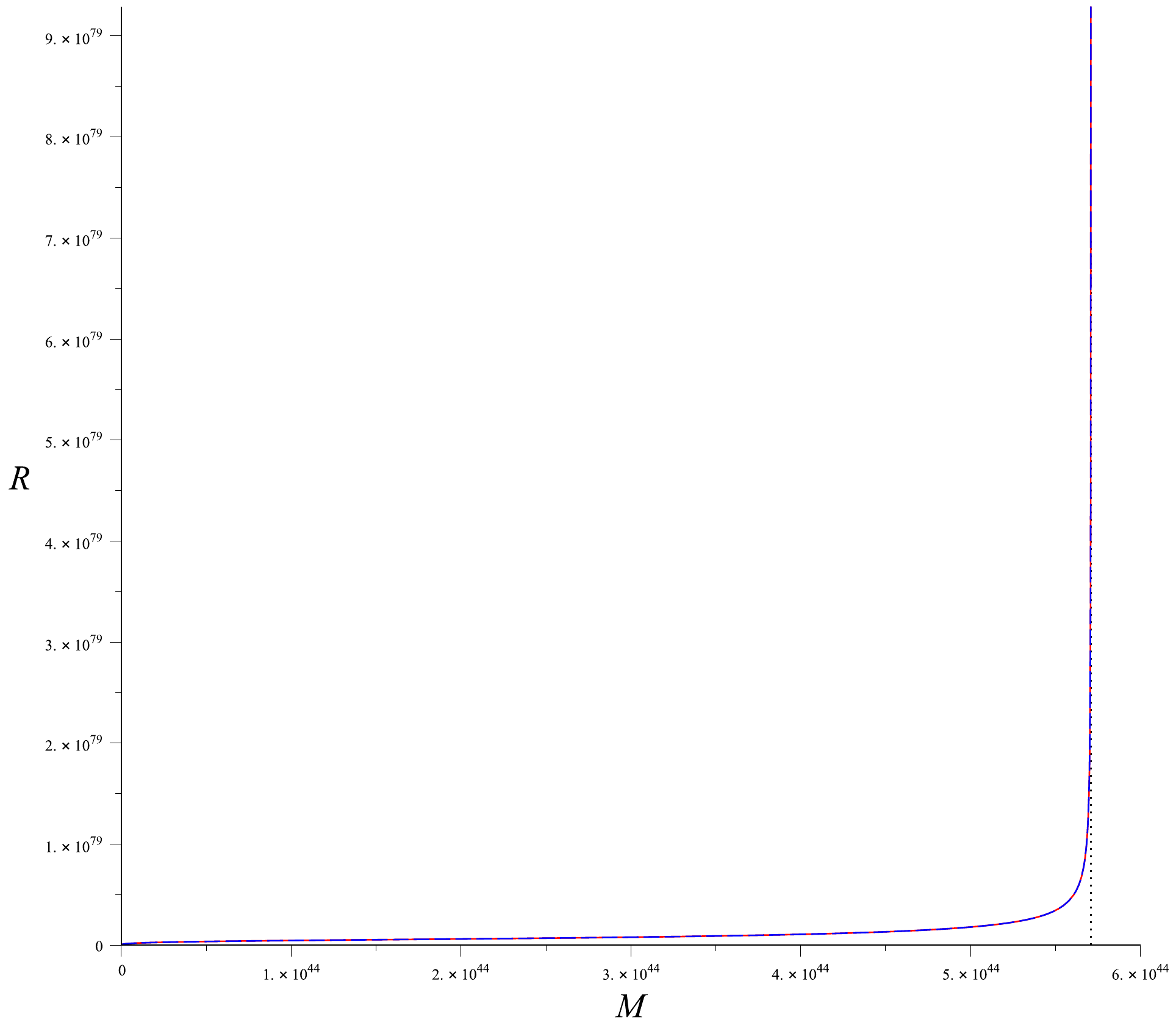}
\caption{The mass-radius relationship of an ultra-relativistic white dwarf with EGUP correction, $R_1(M)$. Without EGUP correction, it is simply the vertical Chandrasekhar limit. The effect of EGUP is to cause sufficiently small white dwarfs to deviate away from the Chandrasekhar limit, but note that no star can exist above the limit. Red curve and blue curve correspond to $\alpha=1$ and $\alpha=-1$ respectively, they pretty much coincide with each other. \label{fig2}} 
\end{figure}

\begin{figure}[!h]
\centering
\includegraphics[width=3.3in]{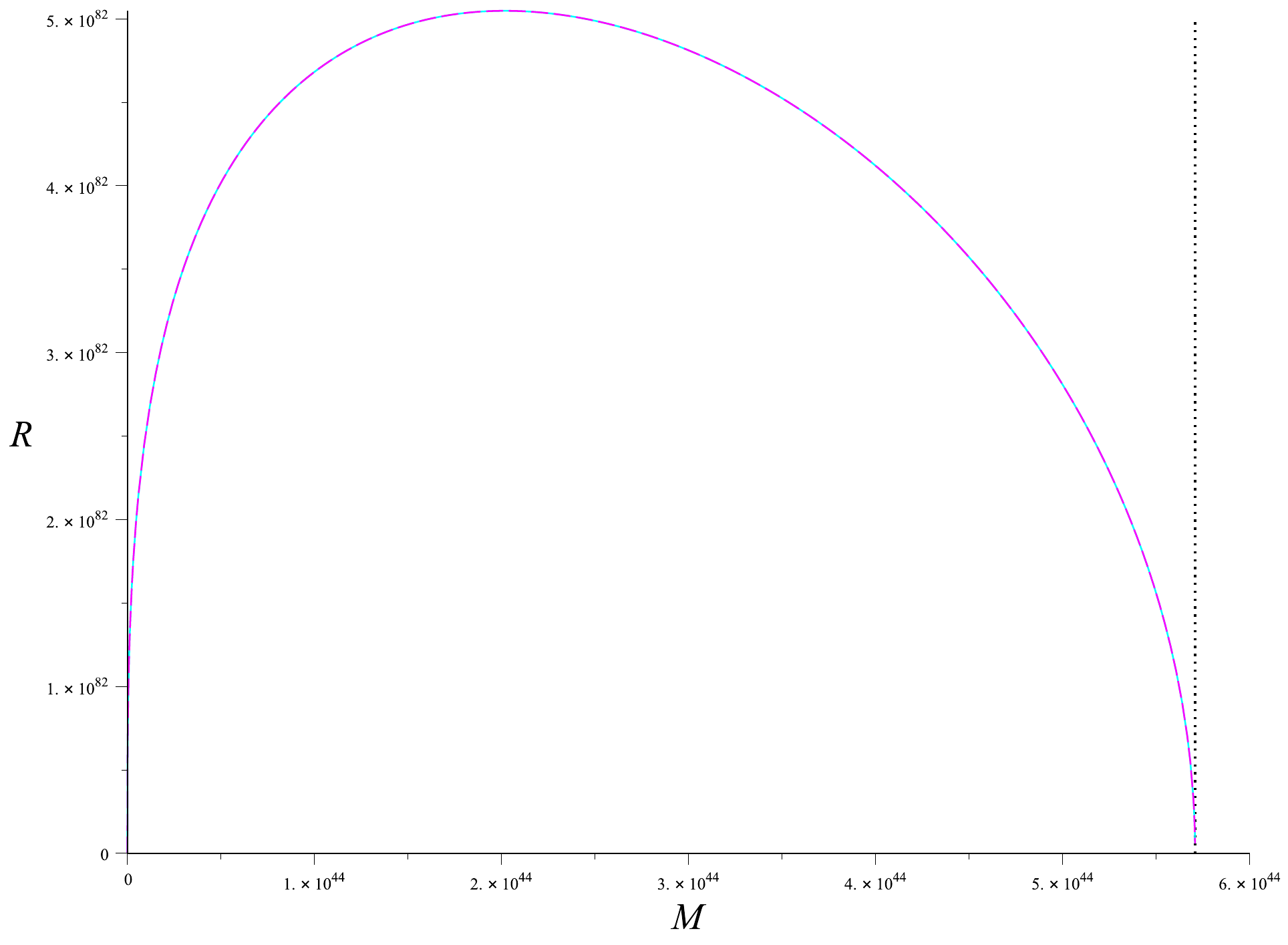}
\caption{The mass-radius relationship of an ultra-relativistic white dwarf with EGUP correction, $R_2(M)$. The cyan curve and magenta curve correspond to $\alpha=1$ and $\alpha=-1$ respectively, but just like for $R_1(M)$, the curves essentially coincide. The curves have a different shape compared to $R_1(M)$, but they are nevertheless bounded by the Chandrasekhar limit. Despite appearance due to the scale involved, the curves do not tend to zero as $M \to M_\text{Ch}^-$, but rather terminate on $M=M_\text{Ch}$ with a finite value. These solutions are likely to be unphysical, see text for discussions. We include them here for completeness. \label{fig3}} 
\end{figure}

For the ultra-relativistic case, again following \cite{1804.05176}, we obtained 
\begin{equation}\label{M-EGUP}
M^{\frac{4}{3}} \sim \frac{c^4R^2 }{2\alpha m_e^{\frac{2}{3}}G^2}\left[1-\sqrt{1-\frac{4G\hbar\alpha}{c^3}\left(\frac{\beta}{L^2} + \frac{M^{\frac{2}{3}}}{m_e^{\frac{2}{3}}R^2}\right)}\right].
\end{equation}
From now onwards, we set $G=c=\hbar=1$.
There are two solutions to Eq.(\ref{M-EGUP}):
\begin{equation}
R_{1,2}(M):=\frac{\sqrt{2}}{2} \left[\frac{L^2M^{\frac{4}{3}}m_e^{\frac{4}{3}}-L^2M^{\frac{2}{3}}\pm \sqrt{\mathcal{F}(\alpha,\beta,M,L)}}{\beta  m_e^{\frac{2}{3}}}\right]^{\frac{1}{2}},
\end{equation}
where
\begin{flalign}\label{F}
\mathcal{F}(\alpha,\beta,M,L)&:=L^4(Mm_e)^{\frac{8}{3}} - 4L^2(Mm_e)^{\frac{8}{3}} \alpha \beta \notag \\ 
&- 2L^4M^2m_e^{\frac{4}{3}} + L^4M^{\frac{4}{3}}.
\end{flalign}
To check whether there is a Chandrasekhar limit, it suffices to plot the functions $R_{1,2}(M)$, which is shown in Fig.(\ref{fig2}) and Fig.(\ref{fig3}). For the plots, we set
$\beta=-3$ and $L=7.31926 \times 10^{60}$. The value of $L$ follows from solving $3/L^2=\Lambda$, taking the ``observed'' $\Lambda=5.6 \times 10^{-122}$ in the Planck units ($1.1056 \times 10^{-52} ~\text{m}^{-2}$ in SI unit). We also take $\alpha=\pm 1$.

One observes that whether the GUP parameter $\alpha$ is negative or positive essentially does not affect the results. This is because in Eq.(\ref{F}), the coefficient for the terms (in the given order) are respectively, of the order $10^{183}, 10^{63}, 10^{213}, 10^{243}$. Thus, the coefficient of the term that contains $\alpha$ is very small compared to the rest. Numerically the sign of $\alpha$, for $\alpha$ not too large (see Discussion), is therefore not important. Both the curves $R_1(M)$ and $R_2(M)$ are bounded by $M=M_\text{Ch}$, so no white dwarfs above the Chandrasekhar limit can exist. $R_2(M)$ behaves drastically different from $R_1(M)$, this means that one could in principle check which solution is physical by comparing with observations, by looking at the masses of ultra-relativistic white dwarfs. 

It does seem that $R_1(M)$ is more physical since it is a ``small deviation'' from the $\alpha=\beta=0$ case. In contrast, $R_2(M)$ develops a high peak for nonzero values of $\alpha$ and $\beta ~(<0)$, no matter how small, but $R_2(M) \equiv 0$ if $\alpha = 0$, so the limit is not smooth. On this ground one might argue that $R_2$ is a spurious, unphysical, solution. Note anyway that $R_2(M_\text{Ch})=({1}/{3})(3^{\frac{3}{4}}\alpha^{\frac{1}{4}}\sqrt{L}/{m_e})$, which is a huge number, despite it seems to go to zero in Fig.(\ref{fig3}), due to the scale involved. What happens is that the curve \emph{terminates} at $M=M_\text{Ch}$.

To prove that neither $R_1$ nor $R_2$ can exceed the Chandrasekhar limit, we first note that\footnote{The term involving $\beta$ is positive, the remaining terms are positive because $(Mm_e)^{\frac{8}{3}} + M^{\frac{4}{3}} \geqslant 2 \sqrt{(Mm_e)^{\frac{8}{3}}M^{\frac{4}{3}}}=2\sqrt{M^4m_e^{8/3}}=2M^2m_e^{4/3}$ by the ``AM-GM'' inequality.} $\mathcal{F} \geqslant 0$. Since the denominator of $R_{1,2}(M)$ is negative ($\because \beta=-3$), in order for $R_1(M)$ to be real, one must have the numerator to be negative as well. This implies that $M^{\frac{4}{3}}m_e^{\frac{4}{3}}-M^{\frac{2}{3}}$ \emph{must} be negative (necessary but not sufficient to guarantee a solution). That is, $M < M_\text{Ch}$. Increasing $L$ has the effect of decreasing the cosmological constant, but this makes $\mathcal{F}$ even larger. Therefore an \emph{arbitrarily small} positive cosmological constant would protect the Chandrasekhar limit. However, the fact that $L$ is \emph{finite} is essential for this to work, since otherwise $R_{1,2}(M) \to 0$ in the limit $L \to \infty$ for fixed $\beta\neq 0$. In fact, we have seen that having GUP-correction only will remove the Chandrasekhar limit. The case for $R_2(M)$ can be similarly argued. 

Finally, we remark on the bound of the cosmological constant. For white dwarfs to exist, we see from Eq.(\ref{nonrelWD}) and Eq.(\ref{M-EGUP}) that we need to impose
\begin{equation}
\Lambda < \frac{M^\frac{2}{3}}{R^2 m_e^{\frac{2}{3}}}  \lesssim \frac{1}{M_\odot^\frac{4}{3} m_e^{\frac{2}{3}}}=10^{-36},
\end{equation}
where in the second inequality, we bound the white dwarf size by the crude estimate $R \sim M$, as indicated by observations\footnote{It also follows from $R > 2M$ of the Schwarzschild limit, relaxed to within $O(1)$ in coefficient in view of possible GUP effect on black hole formation criterion.}, and then using the solar mass as an estimate, $M_\odot \sim 10^{38}$ in Planck units (the electron mass is $m_e \sim 10^{-23}$). The cosmological constant $\Lambda$ is therefore much smaller than the ``natural'' inverse Planck length squared (=1 in Planck unit that we employed). 

Therefore if we accept EGUP as a correct description of Nature, then the existence of white dwarfs in the Universe is consistent with -- in fact it \emph{requires} -- a small cosmological constant, although the bound is still large compared to the observed magnitude of $10^{-122}$. (We assumed the white dwarf to be a pure electron star, but the order of magnitude estimate will not change by much in a realistic white dwarf.)

\section{Discussion}

It was previously found in the literature that GUP with positive parameter $\alpha$ (representing an additional uncertainty due to quantum gravity correction) has the unfortunate effect of removing the Chandrasekhar limit, and therefore seemingly suggests that white dwarfs can be arbitrarily large. Other effects such as black hole formation and realistic astrophysics of stellar structures could prevent this from actually happening. However, it would be more satisfactory to restore the Chandrasekhar limit completely within GUP physics. One way to achieve this is by taking $\alpha$ to be negative, as proposed in \cite{1804.05176}. 

In this work, we show that by considering the so-called extended GUP, an inclusion of an \emph{arbitrarily small but nonzero} cosmological constant protects the Chandrasekhar limit. This is satisfying since our Universe is undergoing an accelerated expansion, with $\Lambda > 0$ being the simplest underlying explanation; and $\Lambda$CDM concordance model remains strong in light of recent observations \cite{1502.01589}. See \cite{0004075, 1306.1527} for reviews and discussions on cosmological constant. There are objections that the smallness of $\Lambda$ implies that it is ``unnatural'' (see, however, \cite{1002.3966}). Indeed the smallness of the cosmological constant is precisely the reason why both signs of $\alpha$ are allowed in our work. It is interesting that GUP removes the Chandrasekhar limit no matter how small $\alpha (>0) $  is, while EGUP restores the limit no matter how small $\Lambda$ is. 

Observationally, white dwarfs are rarely observed to be above the Chandrasekhar limit, although some ``super-Chandrasekhar'' white dwarfs are known to exist \cite{0609616, 1106.3510}. Nevertheless even these rare ones are very close to the Chandrasekhar limit. This is the reason we feel that any modification to the uncertainty principle should not completely remove the Chandrasekhar limit. Although choosing the GUP parameter to be negative achieve this purpose, it nevertheless is not satisfactory for two reasons: firstly, while this is compatible to some quantum gravity scenarios in which physics becomes classical again at the Planck scale, it is incompatible with string theory and general considerations of gravitational correction to quantum uncertainty \cite{9904026}. Secondly, and most importantly, GUP itself has not taken into consideration the fact that our Universe is de Sitter-like, which requires a geometric modification \emph{in addition} to quantum gravitational correction, of the uncertainty principle. While seemingly ad hoc, this is supported by the fact that the uncertainty principle is based on Fourier transform, which is nontrivial on non-flat background. In view of these two reasons, it is satisfying that by considering the effect of cosmological constant one can protect the Chandrasekhar limit (with minor modification), while also allows a wide range of values of both positive and negative GUP parameter.

\begin{figure}[!h]
\centering
\includegraphics[width=3.3in]{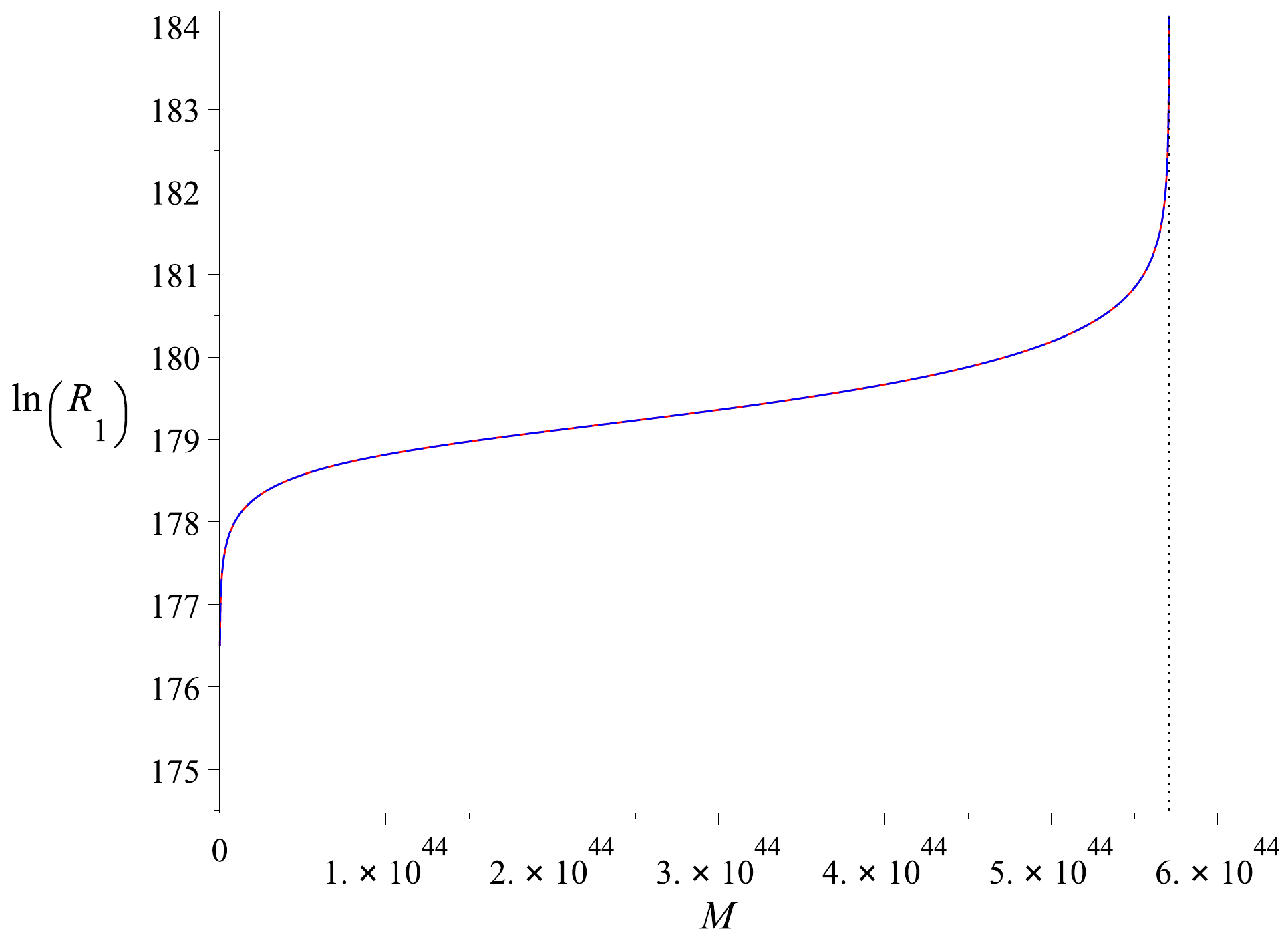}
\caption{The mass-radius relationship of an ultra-relativistic white dwarf with EGUP correction, $\log[R_1(M)]$. Red curve and blue curve correspond to $\alpha=1$ and $\alpha=-1$ respectively, they are still indistinguishable even in log plot. The dashed vertical line corresponds to $M=M_\text{Ch}$.\label{log1}} 
\end{figure}

While the aim of this work is mainly theoretical, let us comment on the observational implications. As mentioned by Moussa in \cite{1512.04337}, the current observation indicates that some white dwarfs have smaller radii than theoretical predictions \cite{0604366,0610073,ironbox}. If we only consider GUP correction, then this favors a negative GUP parameter $\alpha$ since  positive $\alpha$ leads to larger white dwarf for a given mass, not smaller. When we take into account EGUP, whether $\alpha$ is positive or negative does not give noticeable difference when $|\alpha|$ is small. Turning this around, one immediately realizes that the model can be constrained from observations, at least in principle. In Fig.(\ref{log1}), we plotted $\log[R_1(M)]$ as function of white dwarf mass for $|\alpha|=1$, the curves are still not distinguishable.

\begin{figure}[!h]
\centering
\includegraphics[width=3.5in]{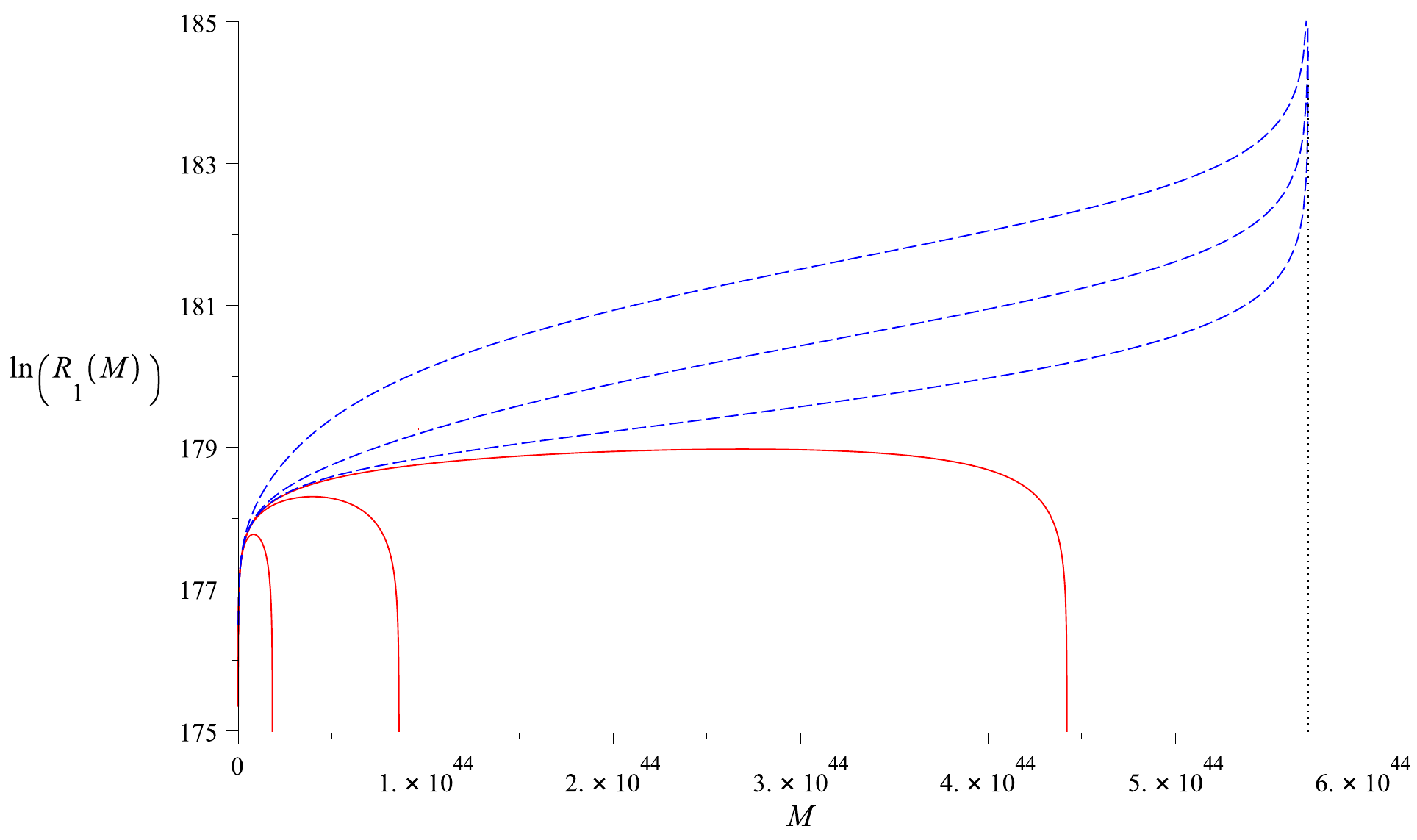}
\caption{The mass-radius relationship of an ultra-relativistic white dwarf with EGUP correction, $\log[R_1(M)]$. 
Solid curves are for $\alpha > 0$ and dashed curves are for $\alpha < 0$.
The curves, from top to bottom, correspond respectively to $\alpha=-10^{113}, -10^{112}, -4\times 10^{110}, 4\times 10^{110}, 10^{112}, 10^{113}$, respectively. 
The dashed vertical line corresponds to $M=M_\text{Ch}$.
\label{log2}} 
\end{figure}

However, as one increases the magnitude of $\alpha$, the curves for positive and negative $\alpha$ gradually separate at around $|\alpha|\sim O(10^{110})$, which are more visible in a log-scale plot, see Fig.(\ref{log2}). The curve corresponds to $\alpha > 0$ now turns around, whereas that of $\alpha < 0$ is an increasing function of the mass.
Similar to asymptotically flat case (i.e. pure GUP case), negative $\alpha$ corresponds to larger white dwarfs for any given mass. In fact, as $|\alpha|$ increases, the curve of $\log[R_1(M, \alpha<0)]$ rises, while the the curve of $\log[R_1(M, \alpha>0)]$  gradually ``shrinks'' towards the left. This is consistent with the known result for GUP, since in that case the curve for $\alpha > 0$ only exist in the region beyond the Chandrasekhar limit \cite{1804.05176}.

Observational data of white dwarfs can therefore in principle be used to constrain the sign of $\alpha$ if $\alpha$ is large enough. In practice there are difficulties: since not all white dwarfs are ultra-relativistic, a more rigorous analysis employing standard Lane-Emden equation will be required to produce the relativistic curves for observational fitting. Even theoretically, $|\alpha| \sim O(10^{110})$ is already questionable, since terrestrial experiments suggest that $\alpha$ is much smaller (e.g., tunneling current measurement yields $\alpha \leqslant 10^{21}$ \cite{0810.5333}), though in such experiments only GUP correction with positive $\alpha$ was considered. 
In other words, it seems likely that our results cannot provide a better constraint on $\alpha$, although it certainly is in agreement with terrestrial experiments. This provides an independent consistency check on GUP physics.  Other white dwarf physics, such as Type Ia supernovae, could potentially yield new constraints in the future. However our model, which assumes a pure electron star, is too crude to investigate such possibilities. 

Finally, we argue that the existence of white dwarfs in our Universe puts an upper bound on the value of $\Lambda$, which we estimated to be $10^{-36}$. The bound is not sharp, and still far larger than the observed value of $10^{-122}$, but it is some 86 order of magnitude improvement compared to the ``natural'' scale of $\Lambda$ expected from quantum field theory. Nevertheless, this improved bound does not explain the origin of the cosmological constant, nor does it explain why the actual observed value is so small \cite{weinberg}, but perhaps it could offer a piece of the puzzle to solve these problems in the future.

\begin{center}
\begin{quote}
``\emph{Its smallness is not petty; on the contrary, it is profound.}'' -- Jan Morris 
\end{quote}
\end{center}

\begin{acknowledgments}
YCO thanks the National Natural Science Foundation of China (grant No.11705162) and the Natural Science Foundation of Jiangsu Province (No.BK20170479) for funding support. 
YCO also thanks Nordita, where part of this work was carried out, for hospitality during his summer visit while participating at the Lambda Program.
The authors thank members of Center for Gravitation and Cosmology (CGC) of Yangzhou University (\href{http://www.cgc-yzu.cn}{http://www.cgc-yzu.cn}) for various discussions and supports.
\end{acknowledgments}

\end{document}